# The Current C, T Transformation Rules of Quantum Field Theory Must Redefine

Mei  Xiaochun

( Department of Physics, Fuzhou University, China)

**Abstract** In light of the $T$ transformation defined in the current quantum field theory, electromagnetic interaction is unchanged under time reversal. However, this kind of time reversal only lets $t \to -t$ in the Hamiltonian of the coordinate space, without considering that the creation and annihilation processes of particles should also be reversed when the concrete problems are calculated in momentum space. In fact, according to the current $T$ transformation of quantum field theory, creation operator of spinor particle is still creation operator and annihilation operator is also still annihilation operator with $Tb_s(\bar{p})T^{-1} = b_s(-\bar{p})$ and $Td_s^+(\bar{p})T^{-1} = d_s^+(-\bar{p})$. This result does not represent the real meaning of time reversal. In the interaction process, a particle's creation operator should become the annihilation operator and its annihilation operator should become the creation operator after time reversal. We should define them as $Tb_s(\bar{p})T^{-1} = b_s^+(-\bar{p})$ and $Td_s^+(\bar{p})T^{-1} = d_s(-\bar{p})$. It is proved that when the reversion of creation and annihilation processes is considered in momentum space, under the condition of high energy, a great symmetry violation of time reversal would be caused in some low order processes of electromagnetic interaction just as the Compton scattering in which the propagation lines of fermions are contained. This result contradicts with the experiments of particle physics and is impossible. Meanwhile, it is proved that the normalization processes of the third order vertex angles of electromagnetic interaction also violate time reversal symmetry, no matter in the coordinate space or in the momentum space. But the symmetry violation is small with a magnitude order about $10^{-5}$. The similar problems exist in the current $C$ transformation of quantum field theory. The $C$ transformation of creation and annihilation operator can not be consistent with the $C$ transformation of spinor particle's wave functions in momentum space. We can obtain the correct $C$ transformation of creation and annihilation operator, but can not obtain the correct $C$ transformation of wave functions $u_s(\bar{p})$ and $v_s(\bar{p})$ in momentum space. Therefore, the current rules of $C,T$ transformations in quantum field theory has serious defect and should be redefined.
**PACC Codes:** 0370, 1130,

## 1. Introduction

In light of the current understanding, the electromagnetic interaction processes of micro-particles is unchanged under time reversal, for the motion equation of quantum mechanics and the interaction Hamiltonian of electromagnetic interaction are invariable under time reversal. On the other hand, as we know that the evolution processes of common macro-material systems which obey the second law of thermodynamics always violate time reversal symmetry. Because macro-systems are composed of atoms and molecules, and atoms and molecules are composed of charged micro-particles, by the logical deductions, the evolution processes of macro-systems should also have the symmetry of time reversal. There exists a basic contradiction here, that is, the so-called irreversibility paradox. Though many theories have been advanced up to now, for example, the theories of coarseness and mixing current and so on[1], none can be regarded to be satisfied. By considering the consistency of nature law, we had to ask whether or not our understanding on the micro-processes has some basic wrong?

In fact, the author's has proved, in the paper tilted "Electromagnetic Retarded Interaction and Symmetry Violation of Time Reversal in Light's High Order Stimulated Radiation and Absorption Processes"[2], that when the retarded effect of radiation fields is taken into account, the light's high order stimulated radiation and stimulated absorption probabilities are not the same so that time reversal symmetry would be violated, though the total Hamiltonian of electromagnetic interaction is still unchanged under time reversal. The reason to cause time reversal symmetry violation is that a certain filial or partial transitive processes of bounding state atoms are forbidden or can't be achieved actually due to the law of energy conservation, the asymmetric actions of effective transition operators before and after time reversal, as well as the special states of atoms themselves. These restrictions would cause the symmetry violation of time reversal of other filial or partial transition processes which can be actualized really. This kind of symmetry violation takes place in the second order processes, so the magnitude order of symmetry violation is quite big when the radiation fields are strong enough. In fact, a great number of experiments have shown that the production processes of laser and most of non-linear optical processes, just as the optical processes of sum frequency, double frequency and different frequency, double stable states[3], self-focusing and self-defocusing[4], echo phenomena[5], as well as optical self-transparence and self observations[6] and so on, are obviously violate the time reversal symmetry. Only because the current theory does not think that time reversal symmetry violation would exists in micro- processes, physicists look at but can not see them. So after the retarded interaction is taken into account, the current formula of light's stimulated radiation and absorption parameters with time reversal symmetry should be revised. A more reliable foundation can be established for the theories of laser and nonlinear optics in which non-equilibrium processes are involved.

Similarly, it is easy to prove that after the retarded electromagnetic interaction is considered, the Hamiltonian of electromagnetic interaction would not keep unchanged under time reversal. The scattering processes of charged micro-particles would also violate the symmetry violation of time reversal. But in this case, the symmetry violation is very small with the magnitude order about $10^{-5}$ direct ratio to $v^3/c^3$. Because the current experimental precision about the experiments of the time reversal is low with the magnitude order about $10^{-2} \sim 10^{-3}$ [7], it is not enough for us to verify such low symmetry violation. We need more accumulate experiments to find the possible symmetry violation in the scattering processes of charged micro- particles. Unfortunately, the existence of the electromagnetic retarded interaction has been neglected for a long time in quantum mechanics. We need to pay attention to this problem when we use the motion equation of quantum mechanics to deal with the problems under extreme conditions such as strong electromagnetic fields, high density and high temperature and so do. In these cases, the electromagnetic retarded effects would become great so that it can not omit.

The result above is based on non-relativity quantum mechanic. In this paper and the follow-up paper "A More Rational and Perfect Scheme of $C,P,T$ Transformations as well as $C,P,T$ Violations in Renormalization Processes of High Order Perturbation in Quantum Field Theory", we discuss the problems of quantum field theory. It is proved that after the high order Perturbation renormalization effects are considered, the electromagnetic interaction processes would violate the symmetry of time reversal. The magnitude order of symmetry violation is also about $10^{-5}$, coincides with that in the non-relative quantum mechanics when the electromagnetic retarded effect is considered. The reason is that the current time reversal definition of particle's creation and annihilation operators are not real time reversal. The $C$ transformation of creation and annihilation operator can not be consistent with the $C$ transformation of spinor particle's wave functions in momentum space. We can obtain the correct $C$ transformation of creation and annihilation operator, but can not obtain the correct $C$ transformation of wave functions $u_s(\bar{p})$ and $v_s(\bar{p})$ in momentum space. By introducing the really correct definitions, $T$ and $C$ symmetry violations would be caused in the high order perturbation renormalization of electromagnetic interaction.

In this way, the origin problem of irreversibility in the evolution processes of macro-systems can be soled well from both quantum mechanics and quantum field theory. The result can also be used to solve the asymmetry origin problem of positive and anti-material in the evolution processes of the universe.

## 2. The problem of T transformation in the current quantum theory of fields

Let's first discuss the problem of time reversal of electromagnetic interaction. In quantum field theory, the $T$ transformations of electromagnetic field $A_\mu$ and spinor field $\psi$ in the coordinate space are defined as individually

$$TA_\mu(\bar{x},t)T^{-1} = -A_\mu(\bar{x},-t) \tag{1}$$

$$T\psi(\bar{x},t)T^{-1} = \tilde{T}\psi(\bar{x},-t) = i\gamma_1\gamma_3\psi(\bar{x},-t) = \sigma_2\psi(\bar{x},-t) \tag{2}$$

$$T\bar{\psi}(\bar{x},t)T^{-1} = [\sigma_2\psi(\bar{x},-t)]^+\gamma_4 = \psi^+(\bar{x},-t)\sigma_2\gamma_4 \tag{3}$$

In which $\sigma_2 \sim \sigma_2 I$, $\sigma_2 = i\gamma_1\gamma_3$ is the Pauli matrix and $I$ is a $2\times 2$ unit matrix. The Hamiltonian of electromagnetic interaction is

$$\mathcal{H}(\bar{x},t) = -\frac{ie}{2}A_\mu(\bar{x},t)\left[\bar{\psi}(\bar{x},t)\gamma_\mu\psi(\bar{x},t) - \psi^\tau(\bar{x},t)\gamma_\mu^\tau\bar{\psi}^\tau(\bar{x},t)\right] \tag{4}$$

In the current quantum field theory, the time reversal of (4) is carried out in light of the following procedure[8]. By considering that the operator $T$ of time reversal is anti-unitary one, we have the time reversal relation $T\alpha T^{-1} = \alpha^*$ and $Ti\gamma_\mu T^{-1} = -i\gamma_\mu^*$. By considering the relations $\sigma_2\gamma_4 = \gamma_4\sigma_2$ and $\sigma_2\gamma_\mu^*\sigma_2 = \gamma_\mu$, based on the definitions (1) ~ (4), we have

$$T\mathcal{H}(\bar{x},t)T^{-1} = -\frac{ie}{2}A_\mu(\bar{x},-t)\left[\psi^+(\bar{x},-t)\sigma_2\gamma_4\gamma_\mu^*\sigma_2\psi(\bar{x},-t) - \psi^\tau(\bar{x},-t)\sigma_2^\tau\gamma_\mu\gamma_4\sigma_2^\tau\psi^{+\tau}(\bar{x},-t)\right]$$

$$= -\frac{ie}{2}A_\mu(\bar{x},-t)\left[\bar{\psi}(\bar{x},-t)\gamma_\mu\psi(\bar{x},-t) - \psi^\tau(\bar{x},-t)\gamma_\mu^\tau\bar{\psi}^\tau(\bar{x},-t)\right] = \mathcal{H}(\bar{x},-t) \tag{5}$$

Because the result is the same to calculate transition probability by using both $\mathcal{H}(\bar{x},-t)$ and $\mathcal{H}(\bar{x},t)$, in this meaning, we say that electromagnetic interaction is unchanged under time reversal.

The transformation above is carried out in the coordinate space and the actions of internal propagation lines of electron and photon are not considered. The Hamiltonian contains all processes of electromagnetic interaction actually such as electron-electron scattering, positive electron-electron scattering, positive electron-electron annihilation, electron-photon scattering and positive electron-photon scattering and so on. The result of (5) indicates that for the sum of all processes, electromagnetic interaction is invariable under time reversal. However, as we known that it is impossible for these processes to take place simultaneously in general. We need to know the transition probability of each single process and its time reversal. The transition probability of a single process is always calculated in the momentum space, in which the internal propagation line of electrons or photons is involved. We take the electron-photon Compton scattering of the second process as an example for further discussion. The process is shown below

$$e^- + \gamma \rightarrow e^- + \gamma$$
$$(p,r) \quad (k,\sigma) \quad (q,s) \quad (l,\rho)$$

For simplification, the interference of identical particles is not considered. By omitting invariable factor, the probability amplitude of process in the momentum space is

$$S \sim -i\bar{u}_s(\bar{q})\varepsilon_\nu^\rho(\bar{l})\gamma_\nu \frac{m - iQ_\alpha\gamma_\alpha}{Q^2 + m^2} \gamma_\mu \varepsilon_\mu^\sigma(\bar{k}) u_r(\bar{p}) \tag{6}$$

Here $Q = p + k$ is the four dimension momentum. According to the current definition, under time reversal, we have $i \rightarrow -i$, $\bar{k} \rightarrow -\bar{k}$, $k_\mu = (\bar{k}, ik_0) \rightarrow -k_\mu$, $\bar{Q} \rightarrow -\bar{Q}$ and $Q_\mu = (\bar{Q}, iQ_0) \rightarrow -Q_\mu$. Therefore, we get

$$T(iQ_\alpha\gamma_\alpha)T^{-1} = -i(-\bar{Q}\cdot\bar{\gamma}, iQ_0\gamma_4)^* = iQ_\alpha\gamma_\alpha^* \tag{7}$$

In light of (2), if we define the time reverse of wave functions in momentum space as

$$Tu_s(\bar{p})T^{-1} = i\gamma_1\gamma_3 u_s(\bar{p}) = \sigma_2 u_s(\bar{p}) \qquad T\bar{u}_s(\bar{p})T^{-1} = u_s^+(\bar{p})\sigma_2\gamma_4 \tag{8}$$

While the time reversal of electromagnetic field in the momentum space is

$$T\varepsilon_\mu^\sigma(\bar{k})T^{-1} = -\varepsilon_\mu^\sigma(\bar{k}) \tag{9}$$

Similar to the transformation of (5), by considering the formulas above and the relations $\sigma_2\gamma_4 = \gamma_4\sigma_2$ as well as $\sigma_2\gamma_\mu^*\sigma_2 = \gamma_\mu$, we obtain the time reversal of (6)

$$S_T = TST^{-1} \sim iu_s^+(\bar{q})\varepsilon_\nu^\rho(\bar{l})\sigma_2\gamma_4\gamma_\nu^* \frac{m - iQ_\alpha\gamma_\alpha^*}{Q^2 + m^2} \gamma_\mu^* \sigma_2 \varepsilon_\mu^\sigma(\bar{k}) u_r(\bar{p})$$

$$= iu_s^+(\bar{q})\gamma_4\varepsilon_\nu^\rho(\bar{l})\sigma_2\gamma_\nu^*\sigma_2 \frac{m - iQ_\alpha\sigma_2\gamma_\alpha^*\sigma_2}{Q^2 + m^2} \sigma_2\gamma_\mu^*\sigma_2\varepsilon_\mu^\sigma(\bar{k}) u_r(\bar{p})$$

$$= i\bar{u}_s(\bar{q})\varepsilon_\nu^\rho(\bar{l})\gamma_\nu \frac{m - iQ_\alpha\gamma_\alpha}{Q^2 + m^2} \gamma_\mu\varepsilon_\mu^\sigma(\bar{k}) u_r(\bar{p}) = -S \tag{10}$$

It indicates $S_T^+ S_T = S^+ S$, i.e., the transition probability is unchanged under time reversal.

This is just the time reversal of the current theory of quantum field. However, it should be pointed out that (11) is not the real time reversal of (6), for it does not contain the reversion of particle's creation and annihilation process which exists actually. As we known in quantum field theory, the process described by (6) indicates that an electron with momentum $\bar{p}$ represented by $u_r(\bar{p})$ and a photon with momentum $\bar{k}$ represented by $\varepsilon_\mu^\sigma(\bar{k})$ are annihilated at the space-time point $x_1$, while an electron with momentum $\bar{q}$ represented by $\bar{u}_s(\bar{q})$ and a photon with momentum $\bar{l}$ represented by $\varepsilon_\nu^\rho(\bar{l})$ are created at the space-time point $x_2$. The time reversal of this process should be that an electron with momentum $-\bar{q}$ represented by $u_s(-\bar{q}) = u_s(\bar{q})$ and a photon with momentum $-\bar{l}$ represented by $-\varepsilon_\nu^\rho(\bar{l})$ are annihilated at the space-time point $x_2$, while an electron with momentum $-\bar{p}$ represented by $\bar{u}_r(-\bar{p}) = \bar{u}_r(\bar{p})$ and a photon with momentum $-\bar{k}$ represented by $-\varepsilon_\mu^\sigma(\bar{k})$ are created at the space-time point $x_1$. But what described by (11) does not represent this process, for it only simply reverses the directions of particle's momentums without reversing the creation and annihilation processes of particles.

In order to see the problem more clearly, let's reword the process to deduce the time reversal of creation and annihilation operators of spinor particles in the current quantum field theory. The quantized spinor field can be written as $\psi(\bar{x},t)=\psi^{(-)}(\bar{x},t)+\psi^{(+)}(\bar{x},t)$ and $\bar{\psi}(\bar{x},t)=\bar{\psi}^{(-)}(\bar{x},t)+\bar{\psi}^{(+)}(\bar{x},t)$ with

$$\psi^{(-)}(\bar{x},t)=\sum_{\bar{p},s}\sqrt{\frac{m}{E}}u_s(\bar{p})b_s(\bar{p})e^{i(\bar{p}.\bar{x}-Et)} \qquad \psi^{(+)}(\bar{x},t)=\sum_{\bar{p},s}\sqrt{\frac{m}{E}}v_s(\bar{p})d_s^+(\bar{p})e^{-i(\bar{p}.\bar{x}-Et)}$$

$$\bar{\psi}^{(-)}(\bar{x},t)=\sum_{\bar{p},s}\sqrt{\frac{m}{E}}\bar{v}_s(\bar{p})d_s(\bar{p})e^{i(\bar{p}.\bar{x}-Et)} \qquad \bar{\psi}^{(+)}(\bar{x},t)=\sum_{\bar{p},s}\sqrt{\frac{m}{E}}\bar{u}_s(\bar{p})b_s^+(\bar{p})e^{-i(\bar{p}.\bar{x}-Et)} \qquad (11)$$

In which $b_s(\bar{p})$ is operator to annihilate a positive spinor particle, $b_s^+(\bar{p})$ is one to create a positive spinor particle. $d_s(\bar{p})$ is one to annihilate a spinor anti-particle and $d_s^+(\bar{p})$ is one to create a spinor anti-particle. $u_s(\bar{p})$ is external factor to annihilate a positive spinor particle in momentum space, $v_s(\bar{p})$ is external factor to create an spinor anti-particle. $\bar{u}_s(\bar{p})$ is external factor to create spinor positive particle and $\bar{v}_s(\bar{p})$ is external factor to annihilate a spinor anti-particle. Because of $\sigma^2=1$, by multiplying $\sigma_2$ on the two sides of (2), we obtain

$$\sigma_2 T\psi(\bar{x},t)T^{-1}=\psi(\bar{x},-t) \qquad (12)$$

According to (11), we have

$$\psi(\bar{x},-t)=\sum_{\bar{p},s}\sqrt{\frac{m}{E}}\left[u_s(\bar{p})b_s(\bar{p})e^{i(\bar{p}.\bar{x}+Et)}+v_s(\bar{p})d_s^+(\bar{p})e^{-i(\bar{p}.\bar{x}+Et)}\right]$$

$$=\sum_{\bar{p},s}\sqrt{\frac{m}{E}}\left[u_s(-\bar{p})b_s(-\bar{p})e^{-i(\bar{p}.\bar{x}Et)}+v_s(-\bar{p})d_s^+(-\bar{p})e^{i(\bar{p}.\bar{x}-Et)}\right] \qquad (13)$$

In the formula, we have considered the fact that the results are the same to sum over $\bar{p}$ and $-\bar{p}$. On the other hand, we relation[8]

$$i\gamma_1\gamma_3 u_s(\bar{p})=\sigma_2 u_s(\bar{p})=u_s^*(-\bar{p}) \qquad i\gamma_1\gamma_3 v_s(\bar{p})=\sigma_2 v_s(\bar{p})=v_s^*(-\bar{p}) \qquad (14)$$

$$\sigma_2 u_s^*(\bar{p})=\sigma^2 u_s(-\bar{p})=u_s(-\bar{p}) \qquad \sigma_2 v_s^*(\bar{p})=\sigma^2 v_s(-\bar{p})=v_s(-\bar{p}) \qquad (15)$$

By considering relation $T\alpha T^{-1}=\alpha^*$ and using (15), we have

$$\sigma_2 T\psi(\bar{x},t)T^{-1}=\sum_{\bar{p},s}\sqrt{\frac{m}{E}}\left[\sigma_2 u_s^*(\bar{p})\ Tb_s(\bar{p})T^{-1}e^{-i(\bar{p}.\bar{x}-Et)}+\sigma_2 v_s^*(\bar{p})\ Td_s^+(\bar{p})T^{-1}e^{i(\bar{p}.\bar{x}-Et)}\right]$$

$$=\sum_{\bar{p},s}\sqrt{\frac{m}{E}}\left[u_s(-\bar{p})\ Tb_s(\bar{p})T^{-1}e^{-i(\bar{p}.\bar{x}-Et)}+v_s(-\bar{p})\ Td_s^+(\bar{p})T^{-1}e^{i(\bar{p}.\bar{x}-Et)}\right] \qquad (16)$$

By comparing (13) with (16), we get

$$Tb_s(\bar{p})T^{-1}=b_s(-\bar{p}) \qquad Td_s^+(\bar{p})T^{-1}=d_s^+(-\bar{p}) \qquad (17)$$

Thus we see that after time reversal, the creation operator of a spinor particle is still a creation operator and the annihilation operator of a spinor particle is also still an annihilation operator, besides the directions of particle's momentums are reversed. Though it seams rational for free particles without creation and annihilation, it is improper for the practical interaction processes with particle's creations and annihilations, in which creation operator should become annihilation operator and annihilation operator should become creation operator after time reversal. So in interaction processes, the rational time reversal transformations of particle's creation and annihilation operators should be

$$Tb_s(\bar{p})T^{-1} = b_s^+(-\bar{p}) \qquad Td_s^+(\bar{p})T^{-1} = d_s(-\bar{p}) \qquad (18)$$

It is obvious that (17) can not describe the time reversal of real processes with interactions.

The same problems also exist for the quantized scalar field and electromagnetic fields based on the time reversal definition (1). Based on it, we deduce the same results that the creation operator of a particle is still a creation operator and the annihilation operator of a particle is also still an annihilation operator after time reversal, but we do not discuss them any more here. So it can be said that there exists basic fault for the definition of time reversal in the current quantum field theory.

In (16), we have tacitly approved the below reversions of spinor fields in momentum space actually

$$Tu_s(\bar{p})T^{-1} = u_s^*(\bar{p}) \qquad Tv_s(\bar{p})T^{-1} = v_s^*(\bar{p}) \qquad (19)$$

From the relations, we have

$$T\bar{u}_s(\bar{p})T^{-1} = [u_s^*(\bar{p})]^\dagger \gamma_4 = u_s^\tau(\bar{p})\gamma_4 \qquad T\bar{v}_s(\bar{p})T^{-1} = [v_s^*(\bar{p})]^\dagger \gamma_4 = v_s^\tau(\bar{p})\gamma_4 \qquad (20)$$

By considering (14), we obtain

$$Tu_s(\bar{p})T^{-1} = u_s^*(\bar{p}) = \sigma_2 u_s(-\bar{p}) \qquad Tv_s(\bar{p})T^{-1} = v_s^*(\bar{p}) = \sigma_2 v_s(-\bar{p}) \qquad (21)$$

On the other hand, the helicity of spinor particle is defined as $s = \bar{\Sigma} \cdot \bar{p}/|\bar{p}|$, in which $\bar{\Sigma}$ is particle's spin. Under time reversal, we have $\vec{\Sigma} \to -\vec{\Sigma}$ and $\bar{p} \to -\bar{p}$, so the helicity $s$ is unchanged. By considering the fact that the wave functions of spinor particles in the momentum space are the eigen states of helicity, when $s$ is unchanged but the directions of particle's momentums are reversed, the forms of wave functions are still unchanged [8]. Speaking concretely, let $\bar{\Sigma} = \bar{\sigma} I$ and $\bar{n} = \bar{p}/|\bar{p}|$, $\bar{\sigma}$ is Pauli matrix and $I$ is 2×2 matrix. The projection of $\bar{\sigma}$ on $\bar{n}$ is $\sigma_n = \bar{\sigma} \cdot \bar{n} = \sigma_1 \sin\theta\cos\varphi + \sigma_2 \sin\theta\sin\varphi + \sigma_3 \cos\theta$. $u_s(\bar{p})$ and $v_s(\bar{p})$ are the function $\theta$ and $\varphi$. If $s$ is unchanged, when $\bar{p} \to -\bar{p}$, we have $\bar{\sigma} \to -\bar{\sigma}$ simultaneously. Therefore, when $s$ is unchanged, when $\bar{p} \to -\bar{p}$, we have

$$u_s(-\bar{p}) = u_s(\bar{p}) \qquad v_s(-\bar{p}) = v_s(\bar{p}) \qquad (22)$$

So (21) is the same with (8).

It is proved below that the reversions of creation and annihilation processes can be reached by using (19). But time reversal symmetry would be violated when we calculate the second processes of electromagnetic interaction in which the internal propagation lines of electrons are contained. For the convenience of calculation, we define matrixes $\bar{\gamma}_\mu$ and $\tilde{\gamma}_\mu$ which are equivalent with $\gamma_\mu$

$$\tilde{\gamma}_\mu = \gamma_4 \gamma_\mu^* \gamma_4 = \gamma_4(-\gamma_1, \gamma_2 - \gamma_3, \gamma_4)\gamma_4 = (\gamma_1, -\gamma_2, \gamma_3, \gamma_4) \qquad (23)$$

$$\bar{\gamma}_\mu = \gamma_4 \gamma_\mu \gamma_4 = (-\gamma_1, -\gamma_2 - \gamma_3, \gamma_4) = (-\bar{\gamma}, \gamma_4) \qquad \tilde{\gamma}_\mu^\tau = (-\gamma_1, -\gamma_2, -\gamma_3, \gamma_4) = \bar{\gamma}_\mu \qquad (24)$$

It can be seen that $\bar{\gamma}_\mu$ and $\tilde{\gamma}_\mu$ are also the Hermitian matrixes with the same communicate relations

$$\bar{\gamma}_\mu \bar{\gamma}_\nu + \bar{\gamma}_\nu \bar{\gamma}_\mu = \tilde{\gamma}_\mu \tilde{\gamma}_\nu + \tilde{\gamma}_\nu \tilde{\gamma}_\mu = \gamma_\mu \gamma_\nu + \gamma_\nu \gamma_\mu = 2\delta_{\mu\nu} \qquad (25)$$

We let

$$\bar{Q}_\mu = \left(-\bar{Q},\ Q_4\right) \quad \text{with} \quad Q_\mu \bar{\gamma}_\mu = \left(-\bar{Q}\cdot\bar{\gamma},\ Q_4\gamma_4\right) = \bar{Q}_\mu \gamma_\mu \qquad (26)$$

According to the definition (19), by using (8), (23) ~ (26), and inserting $\gamma_4^2 = 1$ in the formula, the time reversal of (6) becomes

$$S_T \sim iu_s^\tau(-\bar{q})\varepsilon_\nu^\rho(\bar{l})\gamma_4\gamma_\nu^* \frac{m - iQ_\alpha \gamma_\alpha^*}{Q^2 + m^2} \gamma_\mu^* \varepsilon_\mu^\sigma(\bar{k}) u_r^*(-\bar{p})$$

$$= iu_r^{*\tau}(\bar{p})\varepsilon_\mu^\sigma(\bar{k})\gamma_4\gamma_\mu^+\gamma_4^2 \frac{m - iQ_\alpha \gamma_\alpha^+}{Q^2 + m^2} \gamma_4^2 \gamma_\nu^+ \gamma_4 \varepsilon_\nu^\rho(\bar{l}) u_s(\bar{q})$$

$$= i\bar{u}_r(\bar{p})\varepsilon_\mu^\sigma(\bar{k})\bar{\gamma}_\mu \frac{m - i\bar{Q}_\alpha \gamma_\alpha}{Q^2 + m^2} \bar{\gamma}_\nu \varepsilon_\nu^\rho(\bar{l}) u_s(\bar{q}) \qquad (27)$$

It is obvious that (27) satisfies the demand of creation and annihilation process reversion, so it can represent the time reversal of (6). Meanwhile, by relation $(iQ_\alpha \gamma_\alpha)^+ = -i(\bar{Q}\cdot\bar{\gamma},\ -iQ_0\gamma_4) = iQ_\alpha \bar{\gamma}_\alpha = i\bar{Q}_\alpha \gamma_\alpha$, $(i\bar{Q}_\alpha \gamma_\alpha)^+ = iQ_\alpha \gamma_\alpha$, $\gamma_4 \bar{\gamma}_\alpha \gamma_4 = \gamma_\alpha$ and $\bar{\gamma}_\mu^+ = \bar{\gamma}_\mu$, the complex conjugations of (6) and (27) are individually

$$S^+ \sim iu_r^+(\bar{p})\varepsilon_\mu^\sigma(\bar{k})\gamma_\mu \frac{m - iQ_\alpha \bar{\gamma}_\alpha}{Q^2 + m^2} \gamma_\nu \gamma_4 \varepsilon_\nu^\rho(\bar{l}) u_s(\bar{q})$$

$$= iu_r^+(\bar{p})\varepsilon_\mu^\sigma(\bar{k})\gamma_4^2 \gamma_\mu \gamma_4^2 \frac{m - iQ_\alpha \bar{\gamma}_\alpha}{Q^2 + m^2} \gamma_4^2 \gamma_\nu \gamma_4 \varepsilon_\nu^\rho(\bar{l}) u_s(\bar{q})$$

$$= i\bar{u}_r(\bar{p})\varepsilon_\mu^\sigma(\bar{k})\bar{\gamma}_\mu \frac{m - iQ_\alpha \bar{\gamma}_\alpha}{Q^2 + m^2} \bar{\gamma}_\nu \varepsilon_\nu^\rho(\bar{l}) u_s(\bar{q}) \qquad (28)$$

$$S_T^+ = -iu_s^+(\bar{q})\varepsilon_\nu^\rho(\bar{l})\bar{\gamma}_\nu^+ \frac{m - iQ_\alpha \gamma_\alpha}{Q^2 + m^2} \bar{\gamma}_\mu^+ \varepsilon_\mu^\nu(\bar{k})\ \gamma_4 u_r(\bar{p})$$

$$= -iu_s^+(\bar{q})\varepsilon_\nu^\rho(\bar{l})\gamma_4^2 \bar{\gamma}_\nu \gamma_4^2 \frac{m - iQ_\alpha \gamma_\alpha}{Q^2 + m^2} \gamma_4^2 \bar{\gamma}_\mu \varepsilon_\mu^\nu(\bar{k})\ \gamma_4 u_r(\bar{p})$$

$$= -i\bar{u}_s(\bar{q})\varepsilon_\nu^\rho(\bar{l})\gamma_\nu \frac{m - i\bar{Q}_\alpha \gamma_\alpha}{Q^2 + m^2} \gamma_\mu \varepsilon_\mu^\sigma(\bar{k}) u_r(\bar{p}) \qquad (29)$$

By comparing $S_T$ with $S^+$, as well as $S_T^+$ with $S$, we know that the differences between them are at $Q_\mu$ and $\bar{Q}_\mu$ (or $-\bar{Q}$ and $\bar{Q}$). Only let $Q_\mu \to \bar{Q}_\mu$ (or $\bar{Q} \to -\bar{Q}$) in $S^+ S$, we get $S_T^+ S_T$. In this case, we have $S_T^+ S_T \neq S^+ S$, the transition probability density can not keep unchanged under time reversal.

In order to know the magnitude of symmetry violation, we do the concrete calculation below. The current formula of the Compton scattering is based on the condition that the initial electron is at least with momentum $\bar{p} = 0$. In this case, the calculation can be simplified greatly, but we can not get complete description of symmetry violation. So we calculate the problem under the general situation with $\bar{p} \neq 0$.

Suppose that the electric mass is $m$, the energy of initial state electron is $E_p$, the energy and momentum of initial state photon are $\omega_\kappa$ and $\bar{k}$, the energy and momentum of final state photon are $\omega_l$ and $\bar{l}$ individually. Let

$$\overline{R} = e_\mu^\sigma(\bar{k})\gamma_\mu k'_\alpha \gamma_\alpha e_\nu^\rho(\bar{l})\gamma_\nu = \hat{e}^\sigma(\bar{k})\,\hat{k}'\hat{e}^\rho(\bar{l}) \qquad R = e_\nu^\rho(\bar{l})\gamma_\nu k'_\alpha \gamma_\alpha e_\mu^\sigma(\bar{k})\gamma_\mu = \hat{e}^\rho(\bar{l})\,\hat{k}'\hat{e}^\sigma(\bar{k}) \qquad (30)$$

We take $k'_\alpha = k_\alpha$ before time reversal, and take $k'_\alpha = \bar{k}_\alpha = (-\bar{k},\, ik_0)$ after time reversal in (26). By using the relation $(i\hat{p} + m)u_r(\bar{p}) = 0$, as well as the energy and momentum conservation relation $q = p + k - l$, we can obtain transition probability after the statistical average are carried out over the helicity indexes

$$S^+S \sim \frac{1}{2(Q^2+m^2)^2}\sum_{s=1}^{2}\sum_{r=1}^{2}|\bar{u}_s(q)R u_r(p)|^2$$

$$= \frac{1}{4(p\cdot k)^2}Tr(\hat{p}+im)\overline{R}(\hat{q}+im)R = \frac{1}{4(p\cdot k)^2}(A_1 + A_2) \qquad (31)$$

Here
$$A_1 = Tr\hat{p}\hat{e}^\sigma(\bar{k})\,\hat{k}'\hat{e}^\rho(\bar{l})\hat{p}\hat{e}^\rho(\bar{l})\,\hat{k}\hat{e}^\sigma(\bar{k}) + m^2 Tr\overline{R}R \qquad (32)$$

$$A_2 = Tr\hat{p}\hat{e}^\sigma(\bar{k})\,\hat{k}'\hat{e}^\rho(\bar{l})(\hat{k}-\hat{l})\,\hat{e}^\rho(\bar{l})\hat{k}'\hat{e}^\sigma(\bar{k}) \qquad (33)$$

By using the formula below

$$TrP_1\gamma_{\mu 1}P_2\gamma_{\mu 2}P_3\gamma_{\mu 3}\cdots P_{n-1}\gamma_{\mu\,n-1}P_n\gamma_{\mu\,n} = P_1\cdot P_2 TrP_3\gamma_{\mu 3}\cdots P_{n-1}\gamma_{\mu\,n-1}P_n\gamma_{\mu\,n}$$

$$-P_1\cdot P_3 TrP_2\gamma_{\mu 2}\cdots P_{n-1}\gamma_{\mu\,n-1}P_n\gamma_{\mu\,n}\cdots + P_1\cdot P_n TrP_2\gamma_{\mu 2}\cdots P_{n-1}\gamma_{\mu\,n-1} \quad (n \text{ is an even number}) \quad (34)$$

By the direct calculation, we obtain

$$A_1 = 8(p\cdot k')^2 + 32 p\cdot e^\sigma(\bar{k})\, k'\cdot e^\sigma(\bar{k}) p\cdot e^\rho(\bar{l})\, k'\cdot e^\rho(\bar{l}) - 24 p\cdot k' p\cdot e^\sigma(\bar{k})\, k'\cdot e^\sigma(\bar{k}) \qquad (35)$$

$$A_2 = 8 p\cdot k' k'\cdot (k-l) + 4 k'\cdot (k-l) p\cdot e^\sigma(\bar{k})\, k'\cdot e^\rho(\bar{l})\, e^\sigma(\bar{k})\cdot e^\rho(\bar{l})$$

$$-16 k'\cdot (k-l) p\cdot \hat{e}^\sigma(\bar{k}) k'\cdot e^\sigma(\bar{k}) - 16 k'\cdot (k-l) p\cdot e^\rho(\bar{l})\, k'\cdot e^\rho(\bar{l})$$

$$+32 p\cdot e^\sigma(\bar{k})\, k'\cdot e^\sigma(\bar{k})\, k'\cdot e^\rho(\bar{l})(k-l)\cdot e^\rho(\bar{l}) \qquad (36)$$

The following formulas are used to calculate the statistical average and the sum of photon's polarization

$$\sum_{\sigma=1}^{2}e_\mu^\sigma(\bar{k})e_\nu^\sigma(\bar{k}) = \delta_{\mu\nu} - \frac{1}{\omega_k^2}\left[ k_\mu k_\nu - i\omega_k(k_\mu\delta_{\nu 4} + k_\nu\delta_{\mu 4})\right] \qquad (37)$$

$$\sum_{\rho=1}^{2}e_\mu^\rho(\bar{l})e_\nu^\rho(\bar{l}) = \delta_{\mu\nu} - \frac{1}{\omega_l^2}\left[ l_\mu l_\nu - i\omega_l(l_\mu\delta_{\nu 4} + l_\nu\delta_{\mu 4})\right] \qquad (38)$$

By considering $Q^2 + m^2 = 2(p\cdot k)^2$, we get at last

$$S^+S \sim \frac{1}{8(p\cdot k)^2}\left[ 16(p\cdot k')^2 - 12 p\cdot k' k'\cdot (k-l) - 8 p\cdot k'(B_1 + 4B_2 + 4B_3 + 4B_4)\right.$$

$$+ k' \cdot k (12B_1 - 5B_2 + B_5 + 1) + 32B_1B_2 \Big] \tag{39}$$

Here
$$B_1 = \frac{1}{\omega_k^2} \Big[ p \cdot k \, k' \cdot k + \omega_k (p \cdot k \omega_k + k' \cdot k E_p) \Big]$$

$$B_2 = \frac{1}{\omega_l^2} \Big[ p \cdot l \, k' \cdot l + \omega_l (p \cdot l \omega_l + k' \cdot l E_p) \Big] \tag{40}$$

$$B_3 = \frac{1}{\omega_k^2} \Big\{ k' \cdot l - \omega_k \Big[ k' \cdot k (\omega_k - \omega_l) + k \cdot l \omega_k \Big] \Big\}$$

$$B_4 = \frac{1}{\omega_k^2} \Big[ k' \cdot l \, k \cdot l + \omega_k \omega_l (k \cdot l + k' \cdot l) \Big] \tag{41}$$

$$B_5 = \frac{1}{\omega_k^2 \omega_l^2} \Big\{ p \cdot k k' \cdot l \, k' \cdot l + \omega_l p \cdot k (k' \cdot l \omega_k + k \cdot l \omega_k) + \omega_k \Big[ p \cdot k \, k' \cdot l \omega_l + k \cdot l \, k' \cdot l E_p \Big]$$

$$- \omega_k \omega_l (p \cdot k \, k' \cdot l - p \cdot k \omega_l \omega_k - k' \cdot l \omega_l \omega_k - k \cdot l \omega_l \omega_k) \Big\} \tag{42}$$

By taking $k'_\alpha = k_\alpha$ in (39), we get the transition probability before time reversal. By $k'_\alpha = \bar{k}_\alpha = (-\vec{k}, ik_0)$, we get the transition probability after time reversal. It is obvious that the results are different in both situations with $S^+ S \neq S_T^+ S_T$. For the Compton scattering process, we have to consider the interference effect of identical particles actually, but it does not effect the conclusion of symmetry violation of time reversal so we do not consider it here. Let's estimate the magnitude order of symmetry violation. The first and last items of (39) contain the factor $p^2 \sim m^2$, which is main in the formula under the condition of low energy. We only discuss symmetry violation caused by the first item. The transition probabilities are individually before and after time reversal

$$S^+ S \sim \frac{(\vec{p} \cdot \vec{k} - E_p \omega_k)^2}{(\vec{p} \cdot \vec{k} - E_p \omega_k)^2} \qquad S_T^+ S_T \sim \frac{(\vec{p} \cdot \vec{k} + E_p \omega_k)^2}{(\vec{p} \cdot \vec{k} - E_p \omega_k)^2} \tag{43}$$

So the symmetry violation of time reversal can be described by the formula below

$$\beta = \frac{S_T^+ S_T - S^+ S}{S^+ S} \sim \frac{4 E_p \omega_k \vec{p} \cdot \vec{k}}{(\vec{p} \cdot \vec{k} - E_p \omega_k)^2} \tag{44}$$

Under the condition of low energy with $|\vec{p}| \ll E_p$, we have $\beta \sim 4 \vec{p} \cdot \vec{k} / E_p \omega_k \ll 1$. In this case, the symmetry is very small. If initial electron is at rest with $\vec{p} = 0$ and $E_p = m$, we have $\beta = 0$. The symmetry violation of the first items is zero, though other items still violate symmetry with small magnitude. But under the condition of high energy with $|\vec{p}| \sim E_p$, $\beta$ may be very great so that great symmetry violation would be caused for the second order process of the Compton scattering. However, the experiments of particle physics show that this is impossible for the low order processes to violate the symmetry of time reversal with so big magnitude order.

By the same reason, after the process reversions of particle's creations and annihilations are considered, the other second order processes of electromagnetic interaction to contain the internal propagation lines of electrons such as the scattering between positive electron-photon and the annihilation between positive and negative electrons also violate time reversal symmetry, thought they are actually impossible. But for the other second order processes to contain the internal propagation lines of photons such as the scatterings between electrons as well as positive and negative electrons, there exist no the symmetry violation of time reversal.

### 3. T Violation in the normalization processes of third order vertex angles

In view of a few references reported, the $C,P,T$ transformations in the normalization processes of high order perturbation of electromagnetic interaction are discussed in detail below. When an electron is scattered by external electromagnetic field, by omitting the invariable factor $-e\delta^4(p_2 - p_1 - k)$, the total probability amplitude of the first and third order processes can be written as [10]

$$S \sim -\bar{u}_s(\bar{p}_2)\Gamma_\mu^{(2)}(p_1, p_2) u_r(\bar{p}_1)\varepsilon_\mu^\sigma(\bar{k}) \tag{45}$$

Here $k = p_2 - p_1$. The formula above can be considered as a part of more complex Feynman diagrams. Before regularization and normalization, (45) is symmetric under time reversal according to the current theory. After the regularization is carried out, by separating infinite quantity, we get

$$\Gamma_\mu^{(2)}(p_1, p_2) = (1+L)\gamma_\mu + \Lambda_{f\mu}^{(2)}(p_1, p_2) \sim (1+L)[\gamma_\mu + \Lambda_{f\mu}^{(2)}(p_1, p_2)] \tag{46}$$

In which $L$ is infinite, $\Lambda_{f\mu}^{(2)}(p_1, p_2)$ does not contain ultraviolet divergence but contains infrared divergence. In order to eliminate infrared divergence, we suppose that photon has a small static mass $\rho$ at first. Then let $\rho \to 0$ after finishing calculation. By introducing the charge normalization to let $e \to e(1+L)$, we obtain finite probability amplitude

$$S \sim -\bar{u}_s(\bar{p}_2)[\gamma_\mu + \Lambda_{f\mu}^{(2)}(p_1, p_2)] u_r(\bar{p}_1)\varepsilon_\mu^\sigma(\bar{k}) \tag{47}$$

In which

$$\Lambda_{f\mu}^{(2)}(p_1, p_2) = \gamma_\mu G(p_1, p_2) + K(p_1, p_2)\sigma_{\mu\nu}k_\nu \tag{48}$$

$$\sigma_{\mu\nu} = \frac{1}{2i}(\gamma_\mu\gamma_\nu - \gamma_\nu\gamma_\mu) \tag{49}$$

$$K(p_1, p_2) = \int_0^1 dx \int_0^x dy \frac{1}{q^2} x(1-x)m_0 \tag{50}$$

$$G(p_1, p_2) = \int_0^1 dx \int_0^x dy \left[\int_0^1 dz \frac{1}{q_1^2} k^2 y(x-y) - \frac{1}{q_0^2}(2 - 2x - x^2)m_0\right.$$

$$\left. + \frac{1}{q^2} k^2(1-x+y)(1-y) + \frac{1}{q^2}(2 - 2x - x^2)m_0^2\right] \tag{51}$$

$$q_0^2 = m_0^2 x^2 + \rho^2(1-x) \qquad q_1^2 = q_0^2 + k^2 y(x-y)z \tag{52}$$

$$q^2 = m_0^2 x^2 + p_1^2 x + k(p_1 + p_2)y - (ky + p_1 x)^2 + \rho(1-x) \tag{53}$$

Let $S = S_1 + S_2$, we have

$$S_1 \sim [1+G(p_1,p_2)]\bar{u}_s(\bar{p}_2)\gamma_\mu u_r(\bar{p}_1)\varepsilon_\mu^\sigma(\bar{k}) \tag{54}$$

$$S_2 \sim K(p_1,p_2)\bar{u}_s(\bar{p}_2)\sigma_{\mu\nu}k_\nu u_r(\bar{p}_1)\varepsilon_\mu^\sigma(\bar{k}) \tag{55}$$

According to the current definition of time reversal, we have $i \to -i$, $\bar{k} \to -\bar{k}$, $k_\mu = -k_\mu$ and $p_\mu \to -p_\mu$. So $k^2$, $p^2$ and $k \cdot p$ are the invariable quantities of time reversal. The functions $G(p_1,p_2)$ and $K(p_1,p_2)$ in (54) and (55) are also unchanged under time rsversal. By considering the anti-unitary nature of time reversal operators and the transformations (19), inserting factor $\gamma_4^2 = 1$ in the formula, we have the time reversal of $S_1$

$$S_{1T} \sim -u_s^\tau(\bar{p}_2)\gamma_4\gamma_\mu^*\gamma_4^2 u_r^*(\bar{p}_1)\varepsilon_\mu^\sigma(\bar{k}) = -u_s^\tau(\bar{p}_2)\tilde{\gamma}_\mu\gamma_4 u_r^*(\bar{p}_1)\varepsilon_\mu^\sigma(\bar{k})$$

$$= -u_r^{*\tau}(\bar{p}_1)\gamma_4\tilde{\gamma}_\mu^\tau u_s(\bar{p}_2)\varepsilon_\mu^\sigma(\bar{k}) = -\bar{u}_r(\bar{p}_1)\bar{\gamma}_\mu u_s(\bar{p}_2)\varepsilon_\mu^\sigma(\bar{k}) \tag{56}$$

On the other hand, by taking the complex conjugation of (50) and (52) and inserting $\gamma_4^2 = 1$, we get

$$S_1^+ \sim u_r^+(\bar{p}_1)\gamma_4^2 \gamma_\mu \gamma_4 u_s(\bar{p}_2)\varepsilon_\mu^\sigma(\bar{k}) = \bar{u}_r(\bar{p}_1)\bar{\gamma}_\mu u_s(\bar{p}_2)\varepsilon_\mu^\sigma(\bar{k}) \tag{57}$$

$$S_{1T}^+ \sim -u_s^+(\bar{p}_2)\gamma_4^2 \bar{\gamma}_\mu \gamma_4 u_r(\bar{p}_1)\varepsilon_\mu^\sigma(\bar{k}) = -\bar{u}_s(\bar{p}_2)\gamma_\mu u_r(\bar{p}_1)\varepsilon_\mu^\sigma(\bar{k}) \tag{58}$$

So we have $S_{1T} = -S_1^+$, $S_{1T}^+ = -S_1$ and $S_{1T}^+ S_{1T} = S_1^+ S_1$. Then let's consider the time reversal of $S_2$. Referring to the definition of (45), we define

$$\bar{\sigma}_{\mu\nu} = \gamma_4 \sigma_{\mu\nu} \gamma_4 = \frac{1}{2i}(\bar{\gamma}_\mu \bar{\gamma}_\nu - \bar{\gamma}_\nu \bar{\gamma}_\mu) \qquad \tilde{\sigma}_{\mu\nu} = \frac{1}{2i}(\tilde{\gamma}_\mu \tilde{\gamma}_\nu - \tilde{\gamma}_\nu \tilde{\gamma}_\mu) \tag{59}$$

By considering (24), we have

$$\gamma_4 \sigma_{\mu\nu}^* \gamma_4 = -\frac{1}{2i}(\tilde{\gamma}_\mu \tilde{\gamma}_\nu - \tilde{\gamma}_\nu \tilde{\gamma}_\mu) = -\tilde{\sigma}_{\mu\nu} \tag{60}$$

$$\tilde{\sigma}_{\mu\nu}^\tau = \frac{1}{2i}(\tilde{\gamma}_\mu \tilde{\gamma}_\nu - \tilde{\gamma}_\nu \tilde{\gamma}_\mu)^\tau = -\frac{1}{2i}(\tilde{\gamma}_\mu^\tau \tilde{\gamma}_\nu^\tau - \tilde{\gamma}_\nu^\tau \tilde{\gamma}_\mu^\tau) = -\bar{\sigma}_{\mu\nu} \tag{61}$$

By the relations above and inserting $\gamma_4^2 = 1$ in the formula, the time reversals of $S_2$ and $S$ are individually

$$S_{2T} \sim u_s^\tau(\bar{p}_2)\gamma_4 \sigma_{\mu\nu}^* \gamma_4^2 u_r^*(\bar{p}_1)k_\nu \varepsilon_\mu^\sigma(\bar{k}) = -u_s^\tau(\bar{p}_2)\tilde{\sigma}_{\mu\nu}\gamma_4 u_r^*(\bar{p}_1)k_\nu \varepsilon_\mu^\sigma(\bar{k})$$

$$= -u_r^{*\tau}(\bar{p}_1)\gamma_4 \tilde{\sigma}_{\mu\nu}^\tau u_s(\bar{p}_2)k_\nu \varepsilon_\mu^\sigma(\bar{k}) = \bar{u}_r(\bar{p}_1)\bar{\sigma}_{\mu\nu} u_s(\bar{p}_2)k_\nu \varepsilon_\mu^\sigma(\bar{k}) \tag{62}$$

$$S_T = -\bar{u}_r(\bar{p}_1)\left\{ [1+G(p_1,p_2)]\bar{\gamma}_\mu - K(p_1,p_2)\bar{\sigma}_{\mu\nu}k_\nu \right\} u_s(\bar{p}_2)\varepsilon_\mu^\sigma(\bar{k}) \tag{63}$$

It is easy to prove $\sigma_{\mu\nu}^+ = \sigma_{\mu\nu}$, $\tilde{\sigma}_{\mu\nu}^+ = \tilde{\sigma}_{\mu\nu}$ and $\bar{\sigma}_{\mu\nu}^+ = \bar{\sigma}_{\mu\nu}$. So by taking the complex conjugations of $S_2$ and $S_{2T}$, and inserting $\gamma_4^2 = 1$ in the formula, we obtain

$$S_2^+ \sim u_r^+(\bar{p}_1)\gamma_4^2 \sigma_{\mu\nu}\gamma_4 u_s(\bar{p}_2)k_\nu^* \varepsilon_\mu^\sigma(\bar{k}) = \bar{u}_r(\bar{p}_1)\bar{\sigma}_{\mu\nu} u_s(\bar{p}_2)k_\nu^* \varepsilon_\mu^\sigma(\bar{k}) \tag{64}$$

$$S_{2T}^+ \sim u_s^+(\bar{p}_2)\gamma_4^2 \bar{\sigma}_{\mu\nu}\gamma_4 u_r(\bar{p}_1)k_\nu^* \varepsilon_\mu^\sigma(\bar{k}) = \bar{u}_s(\bar{p}_2)\sigma_{\mu\nu} u_r(\bar{p}_1)k_\nu^* \varepsilon_\mu^\sigma(\bar{k}) \tag{65}$$

Because of $k_\nu^* = (\bar{k},\ -ik_0) = -(-\bar{k},\ ik_0) = -\bar{k}_\nu$, we have at last

$$S_T^+ = -\bar{u}_s(\bar{p}_2)\left\{ [1+G(p_1,p_2)]\,\gamma_\mu - K(p_1,p_2)\sigma_{\mu\nu}k_\nu^* \right\} u_r(\bar{p}_1)\varepsilon_\mu^\sigma(\bar{k})$$

$$= -\bar{u}_s(\bar{p}_2)\left\{ [1+G(p_1,p_2)]\,\gamma_\mu + K(p_1,p_2)\sigma_{\mu\nu}\bar{k}_\nu \right\} u_r(\bar{p}_1)\varepsilon_\mu^\sigma(\bar{k}) \tag{66}$$

$$S = \bar{u}_s(\bar{p}_2)\left\{ [1+G(p_1,p_2)]\,\gamma_\mu + K(p_1,p_2)\sigma_{\mu\nu}k_\nu \right\} u_r(\bar{p}_1)\varepsilon_\mu^\sigma(\bar{k}) \tag{67}$$

$$S^+ = \bar{u}_r(\bar{p}_1)\left\{ [1+G(p_1,p_2)]\,\bar{\gamma}_\mu + K(p_1,p_2)\bar{\sigma}_{\mu\nu}k_\nu^* \right\} u_s(\bar{p}_2)\varepsilon_\mu^\sigma(\bar{k})$$

$$= \bar{u}_r(\bar{p}_1)\left\{ [1+G(p_1,p_2)]\,\bar{\gamma}_\mu - K(p_1,p_2)\bar{\sigma}_{\mu\nu}\bar{k}_\nu \right\} u_s(\bar{p}_2)\varepsilon_\mu^\sigma(\bar{k}) \tag{68}$$

Comparing $S_T$ with $S^+$, as well as $S_T^+$ with $S$, we know that there exists a difference between $\bar{k}_\nu$ and $k_\nu$ (or $-\bar{k}$ and $\bar{k}$). To let $\bar{k}_\nu \to -\bar{k}_\nu$ in $S$, we obtain $S_T^+$. To let $k_\nu^* \to \bar{k}_\nu^*$ (It is also equivalent to let $\bar{k} \to -\bar{k}$) in $S^+$, we obtain $S_T$. That is, to let $k_\nu \to \bar{k}_\nu$ in $S^+ S$, we obtain $S_T S_T^+$. So, similar to the result of the second order electron-photon scattering, we have $S_T^+ S_T \neq S^+ S$ in this case. The transition probability of the normalization process of third order vertex angle violates time reversal symmetry.

We can also prove that the total transition probability of all normalization processes of third order vertex angles in the coordinate space is still unchanged under $T$ transformations. When we use the wave functions of the coordinate space to describe the interaction Hamiltonian, we have the equivalent relations

$$A_\mu(x) \to \varepsilon_\mu^\sigma(\bar{k})e^{ik_\nu x_\nu} \qquad\qquad k_\nu = -i\partial_\nu A_\mu(x)/A_\mu(x) \to -i\partial_\nu \tag{69}$$

So we can always write the total probability amplitude of the third order vertex angle processes as equally

$$S \sim \bar{\psi}_2(\bar{x},t)\left\{ [1+G'(\bar{x},t)]\,\gamma_\mu A_\mu(\bar{x},t) - K'(\bar{x},t)\sigma_{\mu\nu}\partial_\nu A_\mu(\bar{x},t) \right\}\psi_1(\bar{x},t) \tag{70}$$

The formula contains four processes, i.e., the scattering between electron and electron as shown in (34), and the scattering between positive electron and positive electron with $S \sim \bar{v}_s(\bar{p}_2)v_r(\bar{p}_1)a_\mu(k)$, as well as the generations and annihilations processes of electron-positive electron with $S \sim \bar{u}_s(\bar{p}_2)v_r(\bar{p}_1)a_\mu(x)$ and $S \sim \bar{v}_s(\bar{p}_2)u_r(\bar{p}_1)a_\mu(k)$. The operator $\partial_\nu = (\nabla,\,i\partial_t)$ is unchanged under time reversal. And so do for the functions with $G'(\bar{x},t) \to G'(\bar{x},-t)$ and $K'(\bar{x},t) \to K'(\bar{x},-t)$. Under time reversal, we have $i \to -i$ and $A_\mu(\bar{x},t) \to -A_\mu(\bar{x},-t)$. By considering the relation $\sigma_2\gamma_4 = \gamma_4\sigma_2$ and $\sigma_2\sigma_{\mu\nu}^*\sigma_2 = -\sigma_{\mu\nu}$, the time reversal of (69) becomes

$$S_T \sim -\psi_2^+(\bar{x},-t)\left\{ [1+G'(\bar{x},-t)]\sigma_2\gamma_4\gamma_\mu^*\sigma_2 A_\mu(\bar{x},-t) + iK'(\bar{x},-t)\sigma_2\gamma_4\sigma_{\mu\nu}^*\sigma_2\partial_\nu A_\mu(\bar{x},-t) \right\}\psi_1(\bar{x},-t)$$

$$= -\bar{\psi}_2(\bar{x},-t)\left\{ [1+G'(\bar{x},-t)]\,\gamma_\mu A_\mu(\bar{x},-t) - iK'(\bar{x},-t)\sigma_{\mu\nu}\partial_\nu A_\mu(\bar{x},-t) \right\}\psi_1(\bar{x},-t) \tag{71}$$

On the other hand, if let $t \to -t$ in (72) and take $\partial_\nu' = (\nabla,\,i\partial/\partial t)$, we would have

$$S(\bar{x},-t) \sim -\bar{\psi}(\bar{x},-t)\left\{[1+G(\bar{x},-t)]\gamma_\mu A_\mu(\bar{x},-t) - iK(\bar{x},-t)\sigma_{\mu\nu}\partial'_\nu A_\mu(\bar{x},-t)\right\}\psi(\bar{x},-t) \tag{72}$$

Because of $\partial'_\nu \neq \partial_\nu = (\nabla, -i\partial/\partial t)$, we have $S_T(\bar{x},t) \neq S(\bar{x},-t)$. According to the definition of (5), when the interaction Hamiltonian in the coordinate space is unchanged under time reversal, we should have $S_T(\bar{x},t) = S(\bar{x},-t)$. Because this condition can not be satisfied, (73) violates the symmetry of time reversal. That is to say, in light of the current definition of time reversal in quantum field theory, no matter in the momentum space or in the coordinate space, no matter for a single process or for the sum of all processes, the normalization processes of third order vertex angles violate the symmetry of time reversal.

## 4. P, C transformations of the normalization processes of third order vertex angles

Then let's discuss the $P$ transformation of the normalization process of third order vertex angle. According to the current definition, under the $P$ transformation, we have $\bar{x} \to -\bar{x}$, $\bar{k} \to -\bar{k}$, $\bar{p} \to -\bar{p}$, $k_\mu \to (-\bar{k}, ik_0) = \bar{k}_\mu$, $p_\mu \to \bar{p}_\mu$ and $s \to -s$ for helicity. So the quantities $k^2$, $p^2$, $k \cdot p$, $G(p_1, p_2)$ and $K(p_1, p_2)$ are unchanged under the $P$ transformation. In the current quantum field theory, the $P$ transformations of spinor filed and electromagnetic field are defined as individually

$$P\varphi(\bar{x},t)P^{-1} = -\varphi(-\bar{x},t) = -\varphi(\bar{x}) \qquad P\varphi^+(\bar{x},t)P^{-1} = -\varphi^+(-\bar{x},t) = -\varphi^+(\bar{x}) \tag{73}$$

$$P\psi(\bar{x},t)P^{-1} = \gamma_4\psi(-\bar{x},t) = \gamma_4\psi(\bar{x}) \qquad P\bar{\psi}(\bar{x},t)P^{-1} = \bar{\psi}(-\bar{x},t)\gamma_4 = \bar{\psi}(\bar{x})\gamma_4 \tag{74}$$

$$P\vec{A}(\bar{x},t)P^{-1} = -\vec{A}(-\bar{x},t) = -\vec{A}(\bar{x}) \qquad PA_4(\bar{x},t)P^{-1} = A_4(-\bar{x},t) = A_4(\bar{x}) \tag{75}$$

Here $\bar{x} = (-\bar{x},t)$. We can write (74) as $PA_\mu(x)P^{-1} = \bar{A}_\mu(\bar{x})$ for simplification. In the momentum space, we have the $P$ transformation

$$P\varepsilon^\sigma_\mu(\bar{k})P^{-1} \to \left[-\bar{\varepsilon}^\sigma(\bar{k}), \varepsilon^\sigma_4(\bar{k})\right] = \bar{\varepsilon}^\sigma_\mu(\bar{k}) \tag{76}$$

$$Pu_s(\bar{p}_1)P^{-1} = \gamma_4 u_{-s}(-\bar{p}_1) \qquad P\bar{u}_s(\bar{p}_2)P^{-1} = \bar{u}_{-s}(-\bar{p}_2)\gamma_4 \tag{77}$$

By considering relations $\bar{\gamma}_\mu \bar{\varepsilon}^\sigma_\mu(\bar{k}) = \gamma_\mu \varepsilon^\sigma_\mu(\bar{k})$ and $\bar{\sigma}_{\mu\nu}\bar{k}_\nu \bar{\varepsilon}^\sigma_\mu(\bar{k}) = \sigma_{\mu\nu}k_\nu \varepsilon^\sigma_\mu(\bar{k})$, as well as the relations above, we have transformations of $S_1$ and $S_2$

$$S_{1P} \sim \bar{u}_{-s}(-\bar{p}_2)\gamma_4\gamma_\mu\gamma_4 u_{-r}(-\bar{p}_1)\bar{\varepsilon}^\sigma_\mu(\bar{k})$$

$$= \bar{u}_{-s}(\bar{p}_2)\bar{\gamma}_\mu u_{-r}(\bar{p}_1)\bar{\varepsilon}^\sigma_\mu(\bar{k}) = \bar{u}_{-s}(\bar{p}_2)\gamma_\mu u_{-r}(\bar{p}_1)\varepsilon^\sigma_\mu(\bar{k}) \to S_1 \tag{78}$$

$$S_{2P} \sim \bar{u}_{-s}(-\bar{p}_2)\gamma_4\sigma_{\mu\nu}\gamma_4 u_{-r}(-\bar{p}_1)\bar{k}_\nu \bar{\varepsilon}^\sigma_\mu(\bar{k})$$

$$= \bar{u}_{-s}(\bar{p}_2)\bar{\sigma}_{\mu\nu}u_{-r}(\bar{p}_1)\bar{k}_\nu \bar{\varepsilon}^\sigma_\mu(\bar{k}) = \bar{u}_{-s}(\bar{p}_2)\sigma_{\mu\nu}u_{-r}(\bar{p}_1)k_\nu \varepsilon^\sigma_\mu(\bar{k}) \to S_2 \tag{79}$$

It means $S_P \to S$ under the $P$ transformation, so we still have $S_T^+ S_T = S^+ S$. Thought the directions of particle's momentums and helicities are reversed simultaneously, the transition probability is unchanged. It is easy to prove that the process shown in (6) is also unchanged under $P$ transformation. We have

$$S_P \sim -i\bar{u}_{-s}(-\bar{q})\gamma_4\bar{\varepsilon}_\nu^\rho(\bar{l})\gamma_\nu\gamma_4^2 \frac{m-i(-\bar{Q},Q_4)\gamma_\alpha}{Q^2+m^2}\gamma_4^2\gamma_\mu\bar{\varepsilon}_\mu^\sigma(\bar{p})\gamma_4 u_{-r}(-\bar{k})$$

$$= -i\bar{u}_{-s}(\bar{q})\bar{\varepsilon}_\nu^\rho(\bar{l})\bar{\gamma}_\nu \frac{m-i(-\bar{Q},Q_4)\bar{\gamma}_\alpha}{Q^2+m^2}\bar{\gamma}_\mu\bar{\varepsilon}_\mu^\sigma(\bar{p})u_{-r}(\bar{k})$$

$$= -i\bar{u}_{-s}(\bar{q})\varepsilon_\nu^\rho(\bar{l})\gamma_\nu \frac{m-iQ_\alpha\gamma_\alpha}{Q^2+m^2}\gamma_\mu\varepsilon_\mu^\sigma(\bar{p})u_{-r}(\bar{k}) \tag{80}$$

The difference between $S$ and $S_P$ is only at $s \to -s$, but this does not affect the transition probability when we take the sum over the helicity indexes.

At last let's discuss the $C$ transformation of the normalization process of third order vertex angle. In the current quantum field theory, the $C$ transformations of electromagnetic field and spinor filed are defined as

$$CA_\mu(x)C^{-1} = -A_\mu(x) \tag{81}$$

The momentums and helicities of particles and $k_\mu$, are unchanged under $C$ transformation. The $C$ transformation of electromagnetic field in momentum space is

$$C\varepsilon_\mu^\sigma(\bar{k})C^{-1} = -\varepsilon_\mu^\sigma(\bar{k}) \tag{82}$$

The current definition of $C$ transformation of spinor particles are

$$C\psi(x)C^{-1} = \psi_c(x) = \gamma_2\gamma_4\bar{\psi}^\tau(x) = \gamma_2\psi^*(x) \tag{83}$$

$$C^{-1}\bar{\psi}(x)C^{-1} = \bar{\psi}_c(x) = [\gamma_2\psi^*(x)]^+\gamma_4 = \psi^\tau(x)\gamma_2\gamma_4 \tag{84}$$

We now show that (83) would read to inconsistent result, i.e., We can obtain the correct $C$ transformation of creation and annihilation operator, but can not obtain the correct $C$ transformation of wave functions $u_s(\bar{p})$ and $v_s(\bar{p})$ in momentum space. According the (83) and by using relation[8]:

$$\gamma_2 u_s^*(\bar{p}) = v_s(\bar{p}) \qquad \gamma_2 v_s^*(\bar{p}) = u_s(\bar{p}) \tag{85}$$

We have

$$\gamma_2\psi^*(x) = \sum_{\bar{p},s}\sqrt{\frac{m}{E}}\left[\gamma_2 u_s^*(\bar{p})b_s^*(\bar{p})e^{-ip\cdot x} + \gamma_2 v_s^*(\bar{p})(d_s^+)^*(\bar{p})e^{ip\cdot x}\right]$$

$$= \sum_{\bar{p},s}\sqrt{\frac{m}{E}}\left[v_s(\bar{p})b_s^+(\bar{p})e^{-ip\cdot x} + u_s(\bar{p})d_s(\bar{p})e^{ip\cdot x}\right] \tag{86}$$

上式中实际上已令 $b_s^*(\bar{p}) = b_s^+(\bar{p})$，$(d_s^+)^*(\bar{p}) = d_s(\bar{p})$。另一方面，我们又有：

$$\psi_c(x) = C\psi(x)C^{-1} = \sum_{\bar{p},s}\sqrt{\frac{m}{E}}\left[Cu_s(\bar{p})C^{-1}Cb_s(\bar{p})C^{-1}e^{ip\cdot x} + Cv_s(\bar{p})C^{-1}Cd_s^+(\bar{p})C^{-1}e^{-ipx}\right] \tag{87}$$

再按（2.83）式将以上两式进行对照，就得到：

$$Cb_s(\bar{p})C^{-1} = d_s(\bar{p}) \qquad Cd_s^+(\bar{p})C^{-1} = b_s^+(\bar{p}) \tag{88}$$

$$Cu_s(\bar{p})C^{-1} = u_s(\bar{p}) \qquad\qquad Cv_s(\bar{p})C^{-1} = v_s(\bar{p}) \qquad (89)$$

In the $C$ transformation, the annihilation operator $b_s(\bar{p})$ of spinor positive particle exchanges with the annihilation operator $d_s(\bar{p})$ of spinor anti-particle each other and the creation operator $d_s^+(\bar{p})$ of spinor anti-particle exchanges with the creation operator $b_s^+(\bar{p})$ of spinor positive particle. They are correct. But in the $C$ transformation, spinor particle's wave functions in momentum space are unchanged. This is incorrect. According to the rules of Fermanian diagram in quantum field theory, wave function $\bar{u}_s(\bar{p})$ represents to create a spinor positive particle, $u_s(\bar{p})$ represents to annihilate a spinor positive particle, $v_s(\bar{p})$ represents to create a spinor anti-particle, $\bar{v}_s(\bar{p})$ represents to annihilate a spinor anti-particle. The real $C$ transformation should be that $u_s(\bar{p})$ is transformed into $\bar{v}_s(\bar{p})$ and $v_s(\bar{p})$ is transformed into $\bar{u}_s(\bar{p})$ with

$$Cu_s(\bar{p})C^{-1} = \bar{v}_s(\bar{p}) \qquad\qquad Cv_s(\bar{p})C^{-1} = \bar{u}_s(\bar{p}) \qquad (90)$$

In fact, in the concrete calculation of current quantum theory of fields, physicists always use (90). The high order interaction processes of are unchanged under $C$. However, if use (90), we can not have (88). So the current $C$ transformation of quantum theory of fields is inconsistent. Thought in the practical problems, it seems that physicists can always calculate $C$ transformation in the correct manner.

## 5. C, P, T transformations of normalizations of self-energy and vacuum polarization

In order to eliminate the infinite of electron self-energy, the Hamiltonian of electromagnetic interaction should be revised as

$$\mathcal{H} = -ieN(\bar{\psi}\hat{A}\psi) - \delta m \bar{\psi}\psi \qquad (91)$$

The revised Hamiltonian in the coordinate space is still unchanged under $C,P,T$ transformations. But we need to calculate concrete problems in momentum space. For the Compton scattering, after mass renormalization is considered, the total transition probability amplitude of the second and third order processes can be written as [10]

$$S \sim ie^2 \bar{u}_s(\bar{p}_2)\varepsilon_\nu^\rho(\bar{k}_2)\gamma_\nu S_f^{(2)}(p_1 - k_1)\gamma_\mu \varepsilon_\mu^\sigma(\bar{k}_1)u_r(\bar{p}_1) \qquad (92)$$

Here $S_f^{(2)}(p) = (m - i\hat{p})/(p^2 + m^2)$. Let $p = p_1 - k_1$, we have

$$S_f^{(2)}(p) = S_f(p) + S_f(p)\left[\Sigma^{(2)}(p) + i(2\pi)^4 \delta m\right] S_F(p) \qquad (93)$$

$$\Sigma^{(2)}(p) = e^2(2\pi)^8 \int d^4k \gamma_\mu D_f(k) S_f(p-k)\gamma_\mu \qquad (94)$$

Before the calculation of regularization, (101) is unchanged under time reversal with $S^+S = S_T^+ S_T$. Because $\Sigma^{(2)}(p)$ contains infinite, we have to separate it. By regularization calculation, we get

$$\Sigma^{(2)}(p) = -i(2\pi)^4 \delta m + B S_f^{-1}(p) + S_f^{-2}(p)\Sigma_F^{(2)}(p) \qquad (95)$$

Here $B$ is an infinite quantity, but $\Sigma_F^{(2)}(p)$ does not contain ultraviolet divergence again with form

$$\Sigma_f^{(2)}(p) = \frac{ie^2}{(2\pi)^6} \int_0^1 dx \int_0^1 dz \frac{(1-x)\{(i\hat{p}-m)(1-x)[x-2(1+x)z]+m(1+x)\}}{m^2 x^2 + (p^2+m^2)(1-z)xz} \qquad (96)$$

The integral can be written simply as

$$\Sigma_f^{(2)}(p) = i\alpha(p)(p^2+m^2)\frac{i\hat{p}-m}{p^2+m^2} + iF(p) = iA(p)S_f(p) + iF(p) \qquad (97)$$

In which $A(p)$ and $F(p)$ are real number. Substitute (103) and (105) into (100), and let

$$S_f^{(2)}(p) = (1+B)S_F(p) + \Sigma_f^{(2)}(p) \sim (1+B)[S_f(p) + \Sigma_f^{(2)}(p)]$$

$$= (1+B)\left\{[1+iA(p)]S_f(p) + iF(p)\right\} \qquad (98)$$

Then we take charge normalization to let $e \to \sqrt{1+B}\,e$, (100) becomes

$$S \sim ie^2 \bar{u}_s(\bar{p}_2)\varepsilon_v^\rho(\bar{k}_2)\gamma_v\left\{[1+iA(p_1+k_1)]S_f(p_1+k_1) + iF(p_1+k_1)\right\}\gamma_\mu \varepsilon_\mu^\sigma(\bar{k}_1)u_r(\bar{p}_1) \qquad (99)$$

$S_f(p)$, $A(p)$ and $B(p)$ are unchanged under time reversal. The result is similar to the Compton scattering of the second order process, the existence of inertial propagation line of fermions violates time reversal symmetry. This kind of violation has nothing to do with of normalization. It is obvious that the normalization process described by (107) is unchanged under $P,C$ transformations.

The regularization and normalization processes of vacuum polarization are discussed below. The total probability amplitude of vacuum polarization containing a second order process and a fourth order process can be written as [10]

$$S \sim \bar{u}_t(\bar{p}_3)\gamma_\mu u_r(\bar{p}_1)D_{f,\mu\nu}^{(2)}(p_1-p_3)\bar{u}_q(\bar{p}_4)\gamma_\nu u_s(\bar{p}_2) \qquad (100)$$

Here $D_f$ is the factor of photon propagation line. Let $k = p_1 - p_3$, we have

$$D_{f,\mu\nu}^{(2)}(k) = \delta_{\mu\nu}D_f(k) + D_f(k)\Pi_{\mu\nu}^{(2)}(k)D_f(k) \qquad (101)$$

$$\Pi_{\mu\nu}^{(2)}(k) = \delta_{\mu\nu}\left[GD_f^{-1}(k) + \Pi_f^{(2)}(k^2)D_f^{-2}\right] \qquad (102)$$

$$\Pi_f^{(2)}(k^2) = \frac{ie^2}{3(2\pi)^6}\int_0^1 dy \frac{y^2(2y-1)(2y-3)}{m^2+k^2(1-y)} \qquad (103)$$

Here $G$ is an infinite quantity. By the charge renormalization to let $e \to \sqrt{1+G}\,e$, (108) can be written as

$$S \sim \bar{u}_t(\bar{p}_3)\gamma_\mu u_r(\bar{p}_1)\delta_{\mu\nu}\left[1+\Pi_f^{(2)}(k_2)\right]\bar{u}_q(\bar{p}_4)\gamma_\nu u_s(\bar{p}_2) \qquad (104)$$

Because $\Pi_f^{(2)}(k^2)$ is unchanged under $T$ and $P$ transformations with $\bar{k} \to -\bar{k}$, so (104) is invariable. It is obvious that the process is also unchanged under $C$. That is to say that the normalization processes of vacuum polarization are invariable under $C,P,T$ transformations. Therefore, only the symmetry violation of time reversal exists in the normalization processes of high order perturbation of electromagnetic interaction, when we calculate concrete problems in momentum space in which the reversions of particle's creation and annihilation processes are considered.

## 6. Discussion

In the discussion above, we have strictly obeyed the rules of $C,P,T$ transformations defined in quantum field theory. The results show that when we calculate transition probability in the coordinate space, for the sum of all concrete processes, electromagnetic interaction is symmetric under time reversal. This is why we always consider micro-processes with time reversal symmetry up to now. However, when we calculate same concrete processes just as the second order Compton scattering in which the internal propagation lines of fermions are involved, time reversal symmetry would be violated. The reason is that when we calculate the transition probabilities of a certain concrete problem in the momentum space in which particle's creations and annihilations are involved, the reversions of processes would violate the symmetry of time reversal. In the situations of high energy, symmetry violation may be very great. But when we discuss the same problems for the sum of processes in the coordinate space, no reversions of processes are involved, so that no symmetry violation appears. Meanwhile, when we calculate the third order vertex angle problems, no matter in the momentum space or in the coordinate space, no matter for a single process or for the sum of all processes, the normalization processes of the third order vertex angles violate the symmetry of time reversal.

Because the internal propagation lines and the vertex angle are the basic elements in the Feynman diagram of quantum field theory, this kinds of symmetry violations would be general, not only existing in electromagnetic interaction, but also existing in strong and weak interactions, if the rules of time reversal transformation defined in quantum field theory is correct. However, as we known that big symmetry violation of time reversal in the low order processes can not be founded in the experiments. Thought it may be possible to find small symmetry violation of time reversal in the normalization processes of high order perturbation, which would be useful for us to explain the famous problems of the irreversibility origin of macro-systems, big symmetry violation in the low order processes is impossible. Because the processes above are still unchanged under $C,P$ transformations, the results would lead to the $CPT$ violation of low order processes. Of cause, this is unacceptable.

There exists the same problem for the $C$ transformation. For the calculations of concrete problems in the momentum space, the current $C$ transformations are not real ones. We also face the same problem to re-define the $C$ transformation in quantum field theory. So the current $C,P,T$ transformation of quantum field theory has foundational difficulty. We have to look new scheme. These problems will be solved well in the author's another paper "A More Rational and Perfect Scheme of C, P, T Transformations as well as C, P, T Violations in Renormalization Processes of High Order Perturbation in Quantum Field Theory" [11].